\documentclass[prl,twocolumn,showpacs,preprintnumbers,amsmath,amssymb,superscriptaddress]{revtex4-1}
\usepackage{graphicx}
\usepackage{dcolumn}
\usepackage{bm}
\usepackage{subfigure}
\usepackage{color}

\newcommand{\COMMENT}[1]{}

\def\rr{{\bf r}}

\begin{document}

\title{Theory of quasiparticle vortex bound states in Fe-based superconductors: application to LiFeAs}
\author{Y. Wang}
\author{P.J. Hirschfeld}
\affiliation{Department of Physics, University of Florida, Gainesville, Florida 32611, USA}
\author{I. Vekhter}
\affiliation{Department of Physics and Astronomy, Louisiana State University, Baton Rouge, Louisiana 70803-4001, USA}
\date{\today}
\begin{abstract}
Spectroscopy of vortex bound states can provide valuable information on the structure of the superconducting
order parameter. Quasiparticle wavefunctions are expected to leak out in the directions of gap minima or
nodes, if they exist, and scanning tunneling spectroscopy (STS) on these low-energy states should probe the
momentum dependence of the gap. Anisotropy can also arise from  band structure effects, however. We perform a
quasiclassical calculation of the density of states of a single vortex in an anisotropic superconductor, and
show that if the gap itself is not highly anisotropic, the Fermi surface anisotropy dominates, preventing
direct observation of superconducting gap features. This serves as a cautionary message for the analysis of
STS data on the vortex state on Fe-based superconductors, in particular LiFeAs, which we treat explicitly.
\end{abstract}
\pacs{Valid PACS appear here}

\maketitle

{\it Introduction.} Four years after the discovery of the iron-based~\cite{kamihara08,Hsu08} high temperature
superconductors, the structure and symmetry of the gap function are still being debated. There is
considerable experimental evidence that there is no universal gap shape~\cite{Wen11,stewart11,HKMreview},
perhaps in part due to the electronic structure that combines small electron and hole pockets, leading to an
``intrinsic sensitivity"~\cite{Kemper10} to details. It is likely~\cite{HKMreview} that in most cases the gap
has $A_{1g}$ symmetry, which, however, allows a continuous deformation from a full gap to that with nodes on
the Fermi surface (FS) sheets. Bulk experimental probes of the gap structure include specific heat and
thermal conductivity oscillations in an external magnetic field~\cite{VekhterReview,Graser_spht}, performed
on the Fe(Te,Se) system~\cite{Zeng} and P-doped 122 family~\cite{MYamashita:2011} respectively. In both
systems the oscillation pattern was found to be
consistent~\cite{Graser_spht,Vekhter_FeSe,Chubukov_FeSe,MYamashita:2011} with an anisotropic gap with minima
along the $\Gamma-\mathrm{X}$ axis (in the unfolded Brillouin zone), as predicted by spin fluctuation
theories (see, e.g.  Ref.~\onlinecite{HKMreview}).

Order parameter structure is also reflected in the local properties of inhomogeneous superconducting states.
Inhomogeneities may arise due to impurities, and the resulting quasi-bound states in nodal superconductors
have tails that ``leak out" in the nodal directions~\cite{Scalapino_imp_state}, providing a signature of the
amplitude modulation of the gap. The interpretation of these impurity states is complex: disorder potentials
can be of the order of electron volts, and hence relatively high energy processes control the formation of
such states, as well as their contribution to scanning tunneling spectroscopy (STS)
images~\cite{BalatskyZhuVekhterRMP}.

Under an applied magnetic field, inhomogeneous superconductivity arises due to modulation of the order
parameter in a vortex lattice, and bound states localized around the vortex cores appear. In this case,
relevant energy scales are of the order of the gap or lower and the bound states properties are determined by
the shape of the gap and the band features near the Fermi surface. The decay length of the core states is of
order of $\xi_0=v_F/\pi\Delta$, where $v_F$ is the Fermi velocity and $\Delta$ is the gap amplitude.
Consequently, variation of the gap with direction $\mathbf{\widehat{k}}$ at the FS,
$\Delta(\mathbf{\widehat{k}})\neq \text{const}$, directly influences the shape of the core states in real
space, leading to the ``tails'' extending along nodes or minima. Since the decay of these states is
exponential in distance $\rho$ from the center of the vortex (except along true nodes where it follows power
laws), these tails are very clearly seen in local measurements, and can be used to probe the gap
shape~\cite{OFischer:RMP07}. Difficulties of interpretation exist in cuprates, where the coherence length is
short and the cores may nucleate competing order, but in most Fe-based superconductors (FeSC) these
complications are less severe or absent over a wide range of experimentally tunable parameters.

On the other hand, a complex aspect of these latter systems arises due to their multiband nature. The
directional dependence ${\bf v}_F(\mathbf{\widehat{k}})$ also affects the decay length of the core states,
especially when combined with different gap amplitudes on different Fermi surface sheets. In FeSC, the Fermi
surface typically consists of two or three hole pockets and two electron pockets, as represented in the
Brillouin zone corresponding to 1-Fe unit cell. The size and shape of these pockets varies considerably from
family to family. A natural question is whether it is the normal state band structure and the Fermi surface,
or the order parameter shape that determine the salient features of the vortex core states as seen in
experiment, and whether one can draw reliable conclusions about the directions of the gap nodes or minima
based on the real space structure of these states. This is the question we address in the current
Communication.

The competition between the two effects has been explored numerically. For example, the sixfold pattern
observed in 2H-NbSe$_2$ core states \cite{HFHess:1990} can be explained either assuming a weak gap
anisotropy or using the angle-dependent density of states around the Fermi surface~\cite{NHayashi:1996}. In
pnictides it was argued both that the vortex core states are controlled by the order parameter
shape~\cite{XHu:2009} and that the location of the peak in the DOS is determined by the proximity to the band
edge in the electron or hole bands~\cite{DWang:2010}. To gain qualitative insight into this issue we consider
a simple model with both the order parameter and band anisotropy characteristic of the Fe-based
superconductors, and find that in the absence of strong nodes the Fermi velocity anisotropy can dominate the
real-space shape of the vortex core states.

\begin{figure}
\centering
\includegraphics[clip=true,width=0.98\columnwidth]{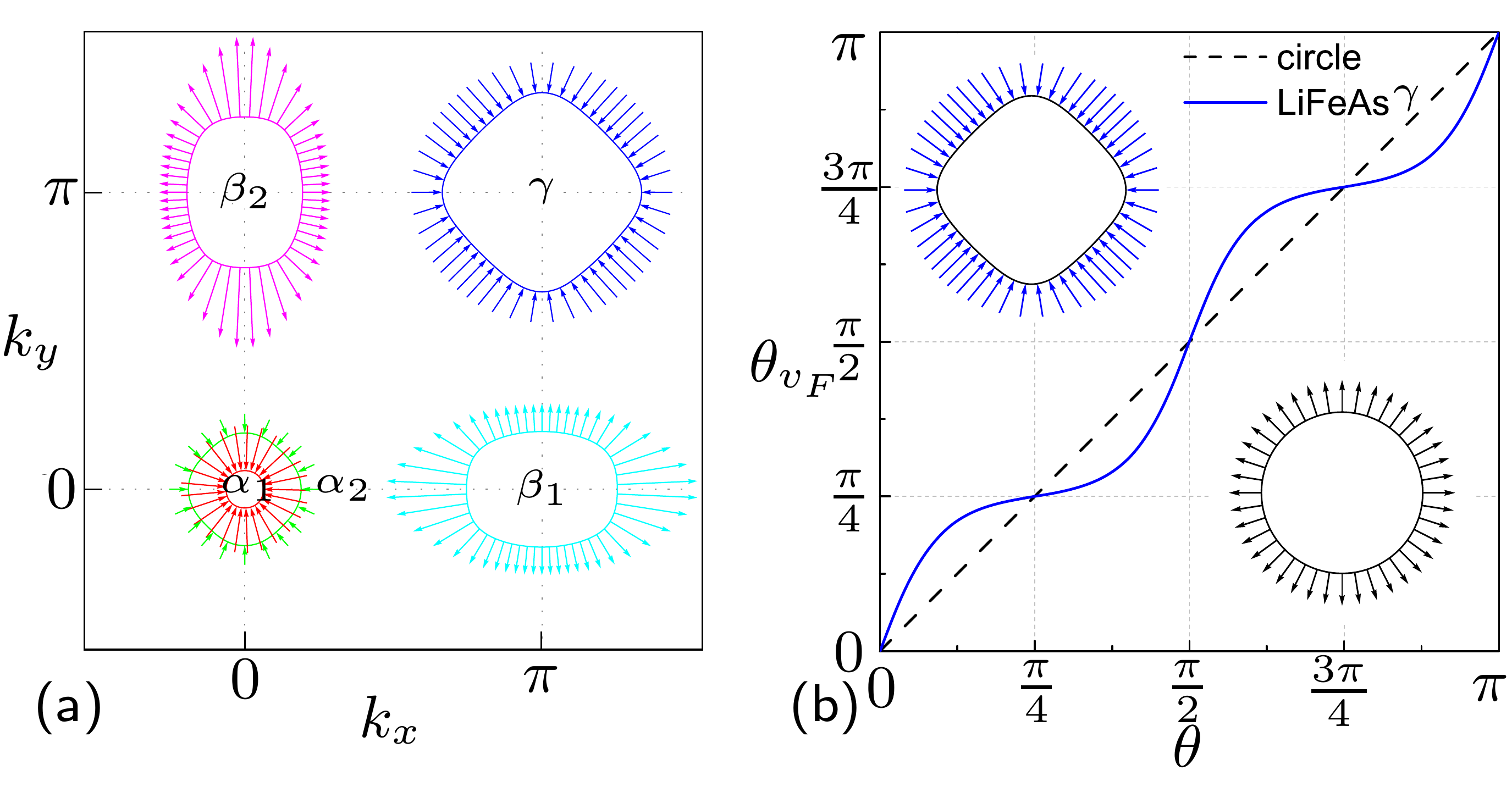}
\caption{(a) Fermi surface of stoichiometric LiFeAs at $k_z=0$ in the unfolded 1-Fe ``effective'' Brillouin
zone from DFT. The Fermi velocities for different sheets are indicated by the arrows pointing to the higher
$E(\mathbf{k})$. We label two inner hole pockets $\alpha_1$, $\alpha_2$, one outer hole pocket $\gamma$ and
two electron pockets $\beta_1$, $\beta_2$. (b) The Fermi velocity direction $\theta_{\mathbf{v}_F}$ vs the
momentum $\mathbf{k}$ azimuthal angle $\theta$ for the LiFeAs $\gamma$ pocket and the circular Fermi surface
(shown as insets).}
 \label{fig:FSvFall}
\end{figure}
We focus on the LiFeAs system, which is ideal for STS measurements due to its nonpolar surfaces. According to
density functional theory (DFT) calculations~\cite{DSingh:2008}, the Fermi surface of this material has three
hole pockets and two electron pockets, see Fig.~\ref{fig:FSvFall}. The outer hole pocket is large and quite
square, according to both DFT results and ARPES~\cite{Borisenko_LiFeAs} and dHvA~\cite{Coldea_LiFeAs}
measurements. Both $\gamma$ and $\alpha_2$ pocket have small Fermi velocities and therefore large normal
state DOS. ARPES has identified superconducting leading edge gaps of order 1.5-2~meV for the hole pockets,
and 3~meV for the electron pockets~\cite{Borisenko_LiFeAs}. The London penetration depth
data~\cite{pendepth_LiFeAs} and specific heat measurements~\cite{SPHT_LiFeAs} ruled out the existence of gap
nodes and were fit to models with two isotropic gaps with
$(\Delta_1,\Delta_2)\simeq(3\,\text{meV},1.5\,\text{meV})$ and $(2\,\text{meV},0.5\,\text{meV})$,
respectively. This suggests moderate gap anisotropy, which is not easily detected by the bulk measurements,
but can substantially affect the real space structure of the core states.

For circular Fermi surfaces the low-energy core bound states extend furthest in the direction of the smallest
gap, but for realistic bands the Fermi velocity anisotropy plays a significant role. Since the cross-sections
of the $\beta_1$ and $\beta_2$ electron pockets rotate by a full 180$^\circ$ along the $k_z$ direction, and
since these gaps are larger, it is unlikely that these sheets contribute substantially to the spatial
anisotropy. We therefore focus on the possible anisotropy of the gap on the hole Fermi surfaces. The most
likely candidate for the anisotropic gap that dominates the low-energy vortex bound states is the $\gamma$
pocket. The orbital content of this pocket is exclusively $d_{xy}$, and it couples only weakly to the
primarily $d_{xz}$ and $d_{yz}$ electron pockets which provide the main pairing weight in the conventional
spin fluctuation approach~\cite{HKMreview}. It is also nearly square, with weakly dispersive parallel
surfaces oriented along the [110] direction in the 1-Fe zone, and with significant variations of the Fermi
velocity between [100] and [110] directions. Hence we first neglect other FS sheets,  and contrast the
results obtained for the $\gamma$ sheet alone with those for a single circular FS.

{\it Model.} We follow the approach of Ref.~\onlinecite{ywang_stewart_11} that relied on the quasiclassical
method for superconductivity~\cite{eilenb68,larkin69,serene83},  used previously to study vortex
cores~\cite{Ichioka97}. The energy-integrated normal and anomalous Green's functions
$g(\mathbf{r},\theta,i\omega_n)$ and $f(\mathbf{r},\theta,i\omega_n)$  obey the coupled Eilenberger equations
\begin{subequations}\label{eq:Eilenberger}
\begin{align}
  &\left[ 2\left( i \omega_n+\frac{e}{c}\mathbf{v}_F\cdot\mathbf{A(r)} \right)+i\hbar\mathbf{v}_F\cdot\nabla
  \right]f(\mathbf{r},\theta,i\omega_n)\notag\\
    &\quad=2ig(\mathbf{r},\theta,i\omega_n)\Delta(\mathbf{r},\theta),\\
  &\left[ 2\left( i \omega_n+\frac{e}{c}\mathbf{v}_F\cdot\mathbf{A(r)} \right)-i\hbar\mathbf{v}_F\cdot\nabla
  \right]\bar{f}(\mathbf{r},\theta,i\omega_n)\notag\\
    &\quad=2ig(\mathbf{r},\theta,i\omega_n)\Delta^{*}(\mathbf{r},\theta),
\end{align}
\end{subequations}
together with the normalization condition $g^2+f\bar{f}=1$. Here $\mathbf{A(r)}$ is the vector potential,
$\mathbf{v}_F$ is the Fermi velocity at the location at the Fermi surface labeled by $\theta$, and
$\omega_n=(2n+1)\pi k_B T$ are fermionic Matsubara frequencies. The Fermi velocity, ${\bf v}_F (\theta)$, is
along the 2D unit vector $\mathbf{\widehat{k}}$ for the circular Fermi surface, and is computed for the
$\gamma$-band in LiFeAs using the Quantum EXPRESSO package\cite{EXPRESSO}, as in Ref. \onlinecite{Graser3D}.
In the low field regime, we consider the problem of an isolated vortex and assume a separable momentum and
coordinate dependence of the order parameter $\Delta(\rho,
\mathbf{\widehat{k}})=\Delta_0\Phi(\theta)\tanh\left(\rho/\eta_r\xi_0\right)$, where $\Delta_0$ is the bulk
gap value in the absence of the field and $\Phi(\theta)$ describes the gap shape on the Fermi surface,
$\Phi_s=1$ , $\Phi_d=\sqrt{2}\cos2\theta$, and $\Phi_{s,ani}=(1-r\cos4\theta)/\sqrt{1+r^2/2}$ with $r=0.3$,
for the isotropic $s$-wave, nodal $d$-wave, and extended $s$-wave gaps respectively. The coherence length is
$\xi_0=\hbar v_{F,\text{rms}}/\Delta_0$ where $v_{F,\text{rms}}=\sqrt{\langle
|\mathbf{v}_F(\widehat{\mathbf{k}})|^2\rangle_{FS}}$, and the brackets denote the normalized average over the
Fermi surface,
\begin{align}
  \langle \cdots \rangle_{FS}=
    \frac{1}{\mathcal{N}}\oint_{FS} \frac{dk_\parallel}{|\mathbf{v}_F(\widehat{\mathbf{k}})|}\cdots
    =\int_0^{2\pi} \frac{d\theta}{2\pi}\tilde{\rho}(\theta)\cdots\,,
\end{align}
where $\mathcal{N}\equiv\oint_{FS} \frac{dk_\parallel}{|\mathbf{v}_F(\widehat{\mathbf{k}})|}$ and
$\tilde{\rho}(\theta)$ is the angle-dependent density of states. The factor $\eta_r$ accounts for the
shrinking of core size at low temperature (Kramer-Pesch effect~\cite{KramerPesch1,KramerPesch2}), and we set
$\eta_r=0.1$ corresponding to $T\sim 0.1T_c$. In a fully self-consistent calculation the gap anisotropy in
momentum space will induce weak core anisotropy in real space~\cite{Ichioka96}, which we ignore here since
the effect is small even for nodal systems~\cite{Ichioka96}.

We solve Eq.~(\ref{eq:Eilenberger}) using the Riccati parameterization~\cite{Dahm02} and integrating along
classical trajectories, $\mathbf{r}(x)=\mathbf{r}_0+x\widehat{\mathbf{v}}_F$ to obtain the functions $g$ and
$f$ at Matsubara frequencies. The LDOS is found after analytic continuation from the retarded propagators,
$N(\rr,\omega)= N_0\langle \mathrm{Re}\,g^R(\mathbf{k}_F,\rr,\omega+i\delta)\rangle_{FS}$. At each point
$\mathbf{r}=(\rho,\phi)$ the LDOS is obtained by summation over the quasiclassical trajectories passing
through $\mathbf{r}$. Each trajectory follows the direction of the Fermi velocity at a given point on the FS,
$\widehat{\mathbf{v}}_F(\widehat{\mathbf{k}})$, and samples the gap
$\Delta(\mathbf{r}(x),\widehat{\mathbf{k}})$. Trajectories sampling regions of small order parameter
contribute to the low energy LDOS. This occurs if the trajectory either passes in the vicinity of the core
where the order parameter is suppressed in real space, $\Delta(\rho)\ll\Delta_0$ (small impact parameter,
dominant for isotropic gaps), or is along the direction where the gap has a node or a deep minimum in
momentum space, $\Delta(\widehat{\mathbf{k}})\ll\Delta_0$ (dominant for nodal superconductivity).

The influence of the FS shape is then clear: the number of trajectories with a given impact parameter depends
on the band structure. Denote the angle between $\widehat{\mathbf{v}}_F$ and $k_x$ axis as
$\theta_{\mathbf{v}_F}$. For a circular FS $\theta_{\mathbf{v}_F}=\theta$, and quasiclassical trajectories in
different directions $\theta_{\mathbf{v}_F}$ are equally weighted in FS averaging. In contrast, for
anisotropic cases, such as the square $\gamma$-sheet in LiFeAs, large parts of the FS have the
${\mathbf{v}_F}$ along the diagonals (see Fig.~\ref{fig:FSvFall}b), and therefore the average over the
trajectories is heavily weighted towards that direction as well.

For an isotropic gap $\Delta(\widehat{\mathbf{k}})=\text{const}$, the largest contribution to the low energy
LDOS at $\mathbf{r}=(\rho,\phi)$ comes from the trajectories passing through the core,
$\theta_{\mathbf{v}_F}=\phi \text{ or } \phi+\pi$. For a cylindrical FS parameterized by angle $\theta$ this
corresponds to two points since $\theta_{\mathbf{v}_F}=\theta$. On an anisotropic FS, such as the $\gamma$
pocket in LiFeAs, many different momentum angles $\theta$ correspond to
$\theta_{\mathbf{v}_F}\approx\pm\frac{\pi}{4}$, and quasiparticles from a large portion of the FS travel
along these directions. For real space direction $\phi=\frac{\pi}{4}$, all these trajectories sample the core
region and contribute to the low energy LDOS. For $\phi$ away from these directions these trajectories have a
nonzero impact parameter and therefore small weight at low energies. For the extended $s$-wave gap model with
$r>0$ in the form factor $\Phi_{s,ani}$, this implies that the regions of large gap will be emphasized due to
preferential directions of $\mathbf{v}_F$, and therefore the FS effects compete with the gap shape in
determining the spatial profile of the vortex core states. Simply assuming that direction of the smallest gap
in $\mathbf{k}$ space yields the orientation of the tails of the bound state wave function need not be
correct, and may be wrong with a strongly anisotropic Fermi surface~\footnote{For a $d$-wave gap along a
circular Fermi surface, near the nodal directions $\theta\approx\frac{\pi}{4}$, the energy spectrum is not
strongly restricted to zero impact parameter. As long as $\theta=\theta_{\mathbf{v}_F}\approx\frac{\pi}{4}$,
the LDOS is enhanced and therefore this case has wider tails along directions $\phi\approx\frac{\pi}{4}$.}.

\begin{figure}[t]
\includegraphics[clip=true,width=0.98\columnwidth]{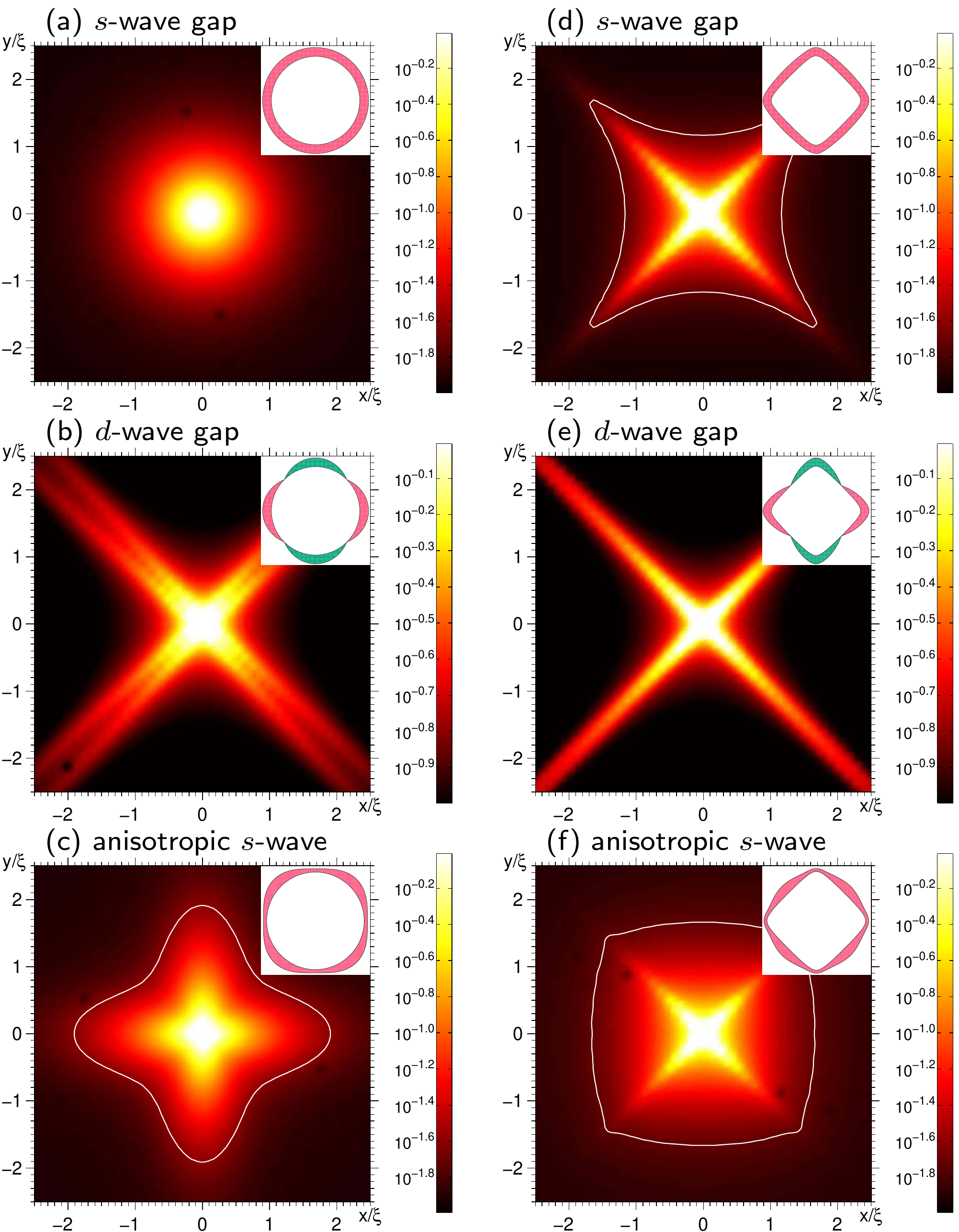}
\caption{Normalized ZDOS in a $2.5\xi_0\times2.5\xi_0$ region around the center of the single vortex for
different gap models with a circular Fermi surface (a -- c) and LiFeAs $\gamma$ pocket (d -- f): (a, d) an
isotropic $s$-wave gap $\Delta_0$; (b, e) a nodal $d$-wave gap $\Delta_0\sqrt{2}\cos2\theta$; (c, f) extended
$s$-wave gap $\Delta_0(1-r\cos4\theta)/\sqrt{1+r^2/2}$, $r=0.3$. The gap bulk value is taken to be
$\Delta_0=1.76T_c$.  The inset on each panel represents a cartoon of the corresponding gap along the Fermi
surface. White contour lines shown correspond to $0.025N_0$.} \label{fig2}
\end{figure}

{\it Results.} Fig.~\ref{fig2} shows the zero energy density of states (ZDOS) of a circular Fermi surface
(a-c) and LiFeAs $\gamma$ pocket (d-f). Comparing panel (a, d) for the isotropic gap, we see that the
rotation symmetry of ZDOS in (a) is broken due to the anisotropy of $\gamma$ pocket and Fermi velocity; at
the same time the ZDOS still preserves the crystal four-fold symmetry. In the $d$-wave case (b) for a
circular Fermi surface, we recover well-known results for the ZDOS, including the double tails along the
nodal directions forced by the vanishing of the bound state wavefunction exactly along the 45$^\circ$
directions in the quasiclassical theory~\cite{Ichioka97}.  While this feature remains, it becomes essentially
invisible in the case of the square Fermi surface shown in panel (e), as the Fermi surfaces concentrates the
quasiparticle trajectories even more in the nodal directions.  Our primary results are now contained in
panels (c) and (f).  The extended-$s$ state  $\Phi_{s,ani}$ has been chosen deliberately to have gap minima
along the $0^\circ$ directions (along the Fe-Fe bond in the FeSC case).  This is clearly visible in the case
of an isotropic pocket (c), as the tails, while not as well-defined as in the true nodal case, extend clearly
along these directions in real space.  These directions rotate by 45$^\circ$, however, when the same gap
exists on the square LiFeAs $\gamma$ pocket, as in (f).  In fact, the ZDOS in panel (f) strongly resembles
the structure observed by Hanaguri et al. in recent STS measurements on LiFeAs~\cite{Hanaguriprivate}.

The results in Fig.~\ref{fig2} strongly challenge the common interpretation of  STS images of vortices, which
assign gap minima to the directions of the extended intensity in real space.  This is probably reasonable in
the case of true nodes, as indicated by the $d$-wave examples shown, but fails if these minima are not
sufficiently deep due to the competition with the Fermi surface effects. Now that the basic structure of this
competition in the case of the  the ZDOS has been understood, it is interesting to ask what may happen in the
case of finite energies $\omega \ne 0$.  The quasiclassical theory incorporates the FS properties solely via
$\mathbf{v}_F$, and thus does not account for the possible changes in the shape of the constant energy
surfaces for STS biases away from zero.  Provided the band shape varies very slowly on the scale of $T_c$,
this neglect should not significantly affect the shape of vortex bound states at nonzero energy, however.  On
the other hand, even within the current model, a more important effect may be included.  In our analysis of
LiFeAs, we have until now neglected all Fermi surface pockets except the outer ($\gamma$) hole pocket, due to
its square shape and because it seems likely to have the smallest gap.  When the bias is increased, higher
energy quasiparticle states, including those associated with larger gaps, will be probed.  Within spin
fluctuation theory~\cite{HKMreview}, both the high density of states $\alpha_2$ pocket, and the electron
pockets, tend to have gap minima along the $0 ^\circ$ directions.  Thus as higher energies are probed, it is
possible that {\it rotations} of the bound state shape may take place as the balance between gap structure
and Fermi surface anisotropy is altered.  Unfortunately even qualitative statements depend on the details of
the sizes of gaps and gap anisotropies on each sheet, as well as on the various Fermi velocities for each
band.  The LiFeAs system is quite clean, however, and if the current controversy between
ARPES~\cite{Borisenko_LiFeAs} and dHvA~\cite{Coldea_LiFeAs} regarding the Fermi surface can be resolved,
spectroscopies of bound states on this system should provide enough information to determine fairly detailed
structure of the gap.

{\it Conclusions.} We have used quasiclassical methods to calculate the vortex bound states within a single
vortex approximation, and highlighted the competition between gap and Fermi surface anisotropy in the
determination of the shape of STS images of vortex bound states. If the Fermi surface anisotropy is large
enough, we have shown that the tails of vortex bound states at low energy need not correspond to the smallest
gaps in the system, if those gaps are not true nodes.  The ZDOS shape measured by STS in experiments on the
LiFeAs system with very clean surfaces is well reproduced by numerical calculation. Within our model, we
attribute the tail-like spectrum to the effect of the non-uniformly distribution of Fermi velocity direction
on the Fermi surface of the LiFeAs $\gamma$ hole pocket. Further measurements of the energy dependence of
bound state shape may further help identify the gap anisotropy.

\begin{acknowledgments}
The authors are grateful to T. Hanaguri and J.C. Davis for useful discussions. YW and PJH were supported by
the DOE under DE-FG02-05ER46236, and  I. V. under
 DE-FG02-08ER46492.
\end{acknowledgments}

\end{document}